 \documentclass[preprint,review,12pt]{elsarticle}

 \usepackage{graphicx}

\usepackage{amssymb}

 \biboptions{comma,sort&compress,numbers}

\journal{Carbon}

\begin{document}

\begin{frontmatter}



\title{Granular superconductivity at room temperature in bulk
highly oriented pyrolytic graphite samples}


\author[]{T. Scheike}
\author[]{P. Esquinazi \fnref{label2}}
\author[]{A. Setzer and W. B\"ohlmann}
\address[]{Division of Superconductivity and Magnetism, Institut
f\"ur Experimentelle Physik II, Universit\"{a}t Leipzig,
Linn\'{e}stra{\ss}e 5, D-04103 Leipzig, Germany}
\fntext[label2]{Corresponding author. Tel/Fax: +49 341 9732751/69.
E-mail address: esquin@physik.uni-leipzig.de (P. Esquinazi)}


\begin{abstract}
We have studied the magnetic response of two bulk highly oriented
pyrolytic graphite (HOPG) samples with different internal
microstructure. For the sample with well defined interfaces,
parallel to the graphene layers, the temperature and magnetic
field hysteresis are similar to those found recently in
water-treated graphite powders. The observed behavior indicates
the existence of granular superconductivity above room temperature
in agreement with previous reports in other graphite samples. The
granular superconductivity behavior is observed only for fields
normal to the embedded interfaces, whereas no relevant hysteresis
in temperature or field is observed for fields applied parallel to
them. Increasing the temperature above $\sim 400$~K changes
irreversibly the hysteretic response of the sample.
\end{abstract}




\end{frontmatter}


\section{Introduction}
\label{intro}

Since the observation of superconductivity in potassium
intercalated graphite $C_8K$\cite{han65},  a considerable amount
of studies reported this phenomenon in graphite based compounds
with critical temperatures $T_c \sim 10$~K, in intercalated
graphite\cite{wel05,eme05}, and above 30~K - though not
percolative - in certain HOPG samples\cite{yakovjltp00,esq08} as
well as in doped graphite
\cite{silvaprl,Kop04,fel09,kopejltp07,han10}. Theoretical work
that deals with superconductivity in graphite and in graphene has
been published in recent years, e.g., $p$-type superconductivity
predicted in inhomogeneous regions of the graphite structure
\cite{gon01} or $d-$wave high-$T_c$ superconductivity\cite{nan12}
based also on resonance valence bonds\cite{doni07}, or at the
graphite surface region due to a topologically protected flat
band\cite{kop11,kop12}. Following a BCS approach in two dimensions
(with anisotropy) critical temperatures $T_c \sim 60~$K have been
estimated if the density of conduction electrons per graphene
plane increases to $n \sim 10^{14}~$cm$^{-2}$, a density that
might be induced by defects and/or hydrogen
ad-atoms\cite{garbcs09} or by Li deposition\cite{pro12}. Further
predictions for superconductivity in graphene support the premise
that $ n > 10^{13}~$cm$^{-2}$ in order to reach $T_c
> 1~$K\cite{uch07,kop08}.

While the existence of defect-induced magnetic order in graphite
reported in ~\cite{yakovjltp00,pabloprb02} was later confirmed  by
independently done studies
\cite{pabloprl03,xia08,ohldagl,zha07,yan09,ohldagnjp,uge10,yaz10,ram10,he11,mia12},
the reproducibility of the phenomenon of high temperature
superconductivity found in some bulk ordered graphite samples
reported previously \cite{yakovjltp00} remain a real challenge. It
has been recently shown that water-treated graphite powder (WTGP)
provides clear signs for granular superconductivity at
temperatures above 300~K \cite{sch12}. In spite of a small
superconducting yield, too small to get any information about the
corresponding superconducting phase(s), the observed behavior
resembled the one expected for a system of Josephson coupled
superconducting grains with critical temperature above room
temperature \cite{sch12}. These last results support earlier
reports on the existence of room-temperature superconductivity in
disordered graphite powders \cite{ant74,ant75} and in HOPG samples
\cite{yakovjltp00,han10}.

Recently performed transport measurements on different graphite
samples of different thickness \cite{bar08,gar12,barjsnm10,sru11}
as well as in transmission electron microscope (TEM) lamellae of HOPG samples \cite{bal12,barint}
suggest that the superconducting regions should be located at some
interfaces between crystalline graphite regions, running parallel
to the graphene planes.
We note that
superconductivity has been found at the interfaces between oxide
insulators \cite{rey07} as well as between metallic and insulating
copper oxides  with  $T_c \gtrsim 50~$K\cite{goz08}. In case of
doped semiconductors the  interfaces
in Bi-bicrystals of inclination type show superconductivity up to
21~K, although Bi bulk is not a superconductor\cite{mun08}.

The aim of this experimental work is then the study of the
magnetic irreversibility (in temperature $T$ as well as in
magnetic field $H$) of HOPG bulk samples and compare it with the
one reported for WTGP. We show that for magnetic fields applied
normal to the internal interfaces  of a HOPG sample the obtained
behavior is similar to that found in WTGP but  with some
interesting differences. Increasing the temperature of the sample
above 400~K changes the hysteretic response in an irreversible
way. For a second HOPG sample without those interfaces we did not
find any relevant irreversibility in $T$ or $H$. Our results
support the view that the superconducting regions are quasi-two
dimensional and localized very likely at the embedded interfaces,
recently found to exist in some HOPG samples \cite{bar08,gar12},
running parallel to the graphene planes of the graphite structure.

\section{Experimental details and samples characteristics}

Magnetic moment measurements of two HOPG samples from different
batches, both nominally ZYA grade (Advanced Ceramics, now
Momentive Performance Materials Inc.), were performed using a
commercial MPMS-7 SQUID magnetometer. The magnetic impurity
concentration determined by PIXE for these samples is below 1~ppm,
see for example ~\cite{jems08,ohldagnjp}, with excellent
reproducibility for all samples of the same grade. Before the
magnetic measurements were done, the samples surface layers were
removed with scotch tape and afterwards they were cleaned in an
ultrasonic bath. The masses of the samples and size (length
$\times$ width $\times$ thickness) were: $m \simeq 1.3~$mg~($\sim
3 \times 2 \times 1~$mm$^3$) for the HOPG-1 and $\simeq 2.3$~mg
($\sim 4 \times 3 \times 0.8~$mm$^3$) for the HOPG-2 sample. Each
sample was fixed on the middle of a long quartz glass stick,
especially designed to reduce to a minimum any magnetic
contribution from the sample holder. The magnetic moment
measurements were performed with the applied magnetic field
parallel to the $c$-axis of the HOPG samples. A few measurements
were performed also with the field parallel to the graphene planes
to check for the anisotropic response of the signals, as was found
in ~\cite{yakovjltp00}. Hysteresis  were measured at different
temperatures in zero-field-cooled (ZFC) and field-cooled (FC)
states at different applied fields as well as a function of field
at constant temperatures. For temperatures above 390~K an
available SQUID oven was used.

Although both HOPG samples show an apparent similar Bernal ordered
structure as well as  similar diamagnetic background signals, we
show below that the observed irreversibilities in field and
temperature are related to the internal structure of the samples,
in particular to the existence of the internal interfaces, as
proposed recently in ~\cite{bar08,barjsnm10,sru11,gar12,ana12}.
The results obtained for the studied samples rule out the
influence of unknown artifacts of the SQUID as well as of the used
background subtraction procedure.

\begin{figure}[]
\begin{center}
\includegraphics[width=.8\columnwidth]{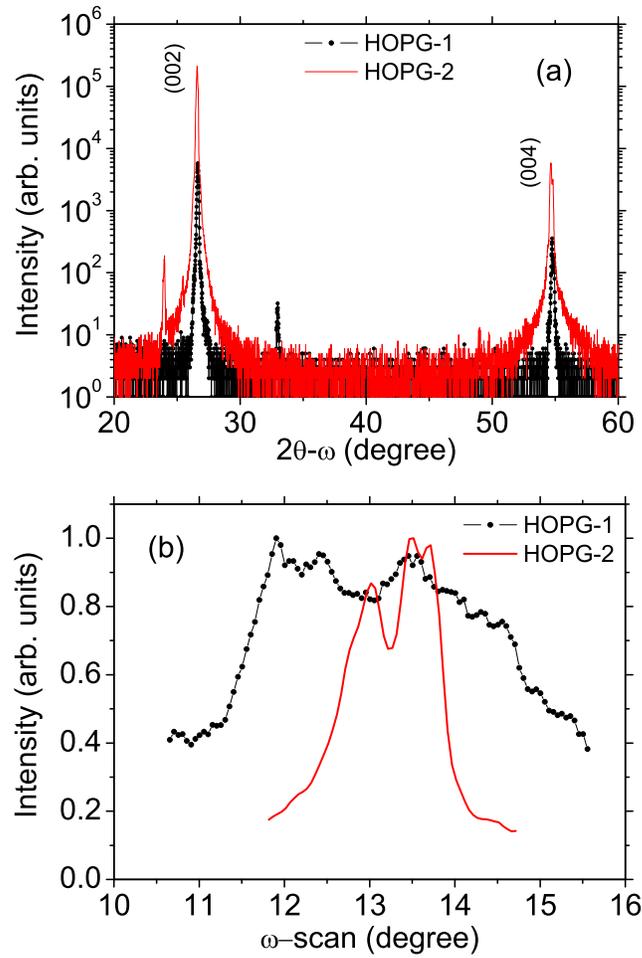}
\caption[]{(a) Wide angle diffractrogram of the two HOPG samples.
(b) Rocking curves of the two HOPG samples at the (002) Bragg
peak.} \label{xr}
\end{center}
\end{figure}

In general the main structural phase of commercial HOPG samples is
the Bernal hexagonal one with the AB... stacking sequence along
the $c-$axis. Taking into account theoretical predictions of the
existence of high temperature surface superconductivity in
rhombohedral graphite \cite{kop12}, it is of interest to check
whether the investigated samples, specially the HOPG-1 one, shows
some evidence for isolated rhombohedral crystallites. X-rays
measurements were done in the two selected HOPG samples with a
Philips X'Pert diffractometer. In the wide-angle diffractogram,
see Fig.~\ref{xr}(a), only the (002) and (004) reflections are
visible. From the Bragg (002) maximum at  $2 \theta = 26.60^\circ$
we estimate a lattice constant $c = 6.70~\AA$. Because the Bragg
angle for the [001] family of reflexes for the hexagonal (Bernal)
and for the rhombohedral graphite structures is the same, the
observed Bragg reflexes do not provide any information about the
possible existence of rhombohedral crystallites. According to
~\cite{lin12} diffraction peaks at 43.45$^\circ$ and 46.32$^\circ$
were found in graphite bulk samples, which can be ascribed to
(10-11) and (10-12) reflections of rhombohedral phase in
coexistence within  the hexagonal crystalline phase matrix. For
our samples and within experimental resolution no diffraction
peaks at those angles are observed indicating that the amount of
rhombohedral phase should be much smaller than 1\%.

A clear difference in the internal structure of the two HOPG
samples can be observed via the rocking curve measurement. Rocking
curves are primarily used to study defects such as dislocation
density, mosaic spread, curvature, misorientation and
inhomogeneity. A comparison of the rocking curves of samples
HOPG-1 and HOPG-2, see Fig.~\ref{xr}(b) clearly indicates that the
sample HOPG-1 has a broader rocking curve width. This broadening
can come from defects like mosaicity, dislocations, and
disruptions in the perfect parallelism of atomic planes, pointing
that sample HOPG-1 has a larger ``disorder" than the HOPG-2
sample.

\begin{figure*}[]
\begin{center}
\includegraphics[width=0.8\columnwidth,angle=+90]{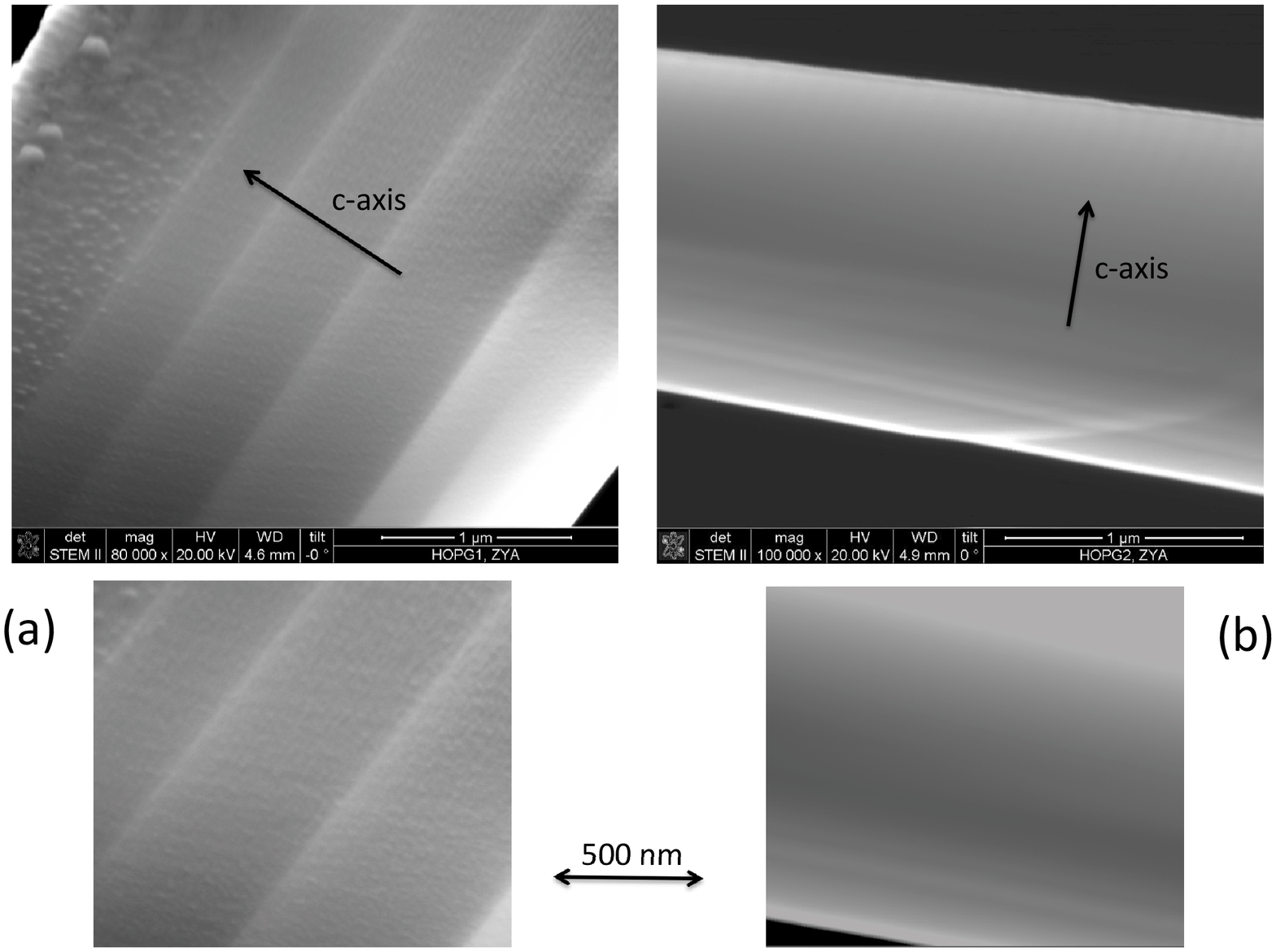}
\caption[]{Low energy transmission electron microscope pictures of
regions of two lamellae of equal thickness (300~nm) of samples
HOPG-1 (a) and HOPG-2 (b) at two different resolutions.}
\label{TEM}
\end{center}
\end{figure*}

Low-energy (20 keV) TEM pictures of the internal microstructure of
the two samples are shown in Fig.~\ref{TEM}. We have used the same
procedure as described in ~\cite{barnano10,dus11,ana12} to
minimize  the influence of Ga$^+$ ions on the graphite structure
during cutting of the lamella. It is interesting to note that
already by the cutting of the lamellae using a dual beam
microscope (FEI Nanolab XT200) we realized that there was a clear
difference between the two samples. Whereas the TEM lamellae of
the HOPG-1 sample could be obtained without any special
difficulties, this was not the case for the HOPG-2 sample. The
HOPG-2 sample got slightly electrically  charged during the
cutting process. A possible reason for the different behavior
could be due to the parallel, high conductivity  of the interfaces
\cite{gar12} (or graphene layers with certain lattice defects)
that exist in the HOPG-1 sample only. We note that  the earlier
observed metallic conductivity of graphite is not intrinsic but
mainly due to  well defined interfaces \cite{bar08,gar12,ana12},
or lattice defects \cite{arn09}.

  A comparison between the TEM pictures obtained for
the two samples clearly indicate the existence of  interfaces in
the HOPG-1 sample. In contrast, the HOPG-2 sample appears much
more homogeneous, without clear interface regions, see
Fig.~\ref{TEM}. The pictures suggest that the difference in the
lattice ``disorder" observed in X-ray diffraction between the two
samples could be related to the existence of a different mosaicity
and therefore to the well defined embedded interfaces between
crystalline regions of the graphite structure. All these results
together with magnetization ones discussed below suggest that the
superconducting phase(s) should be  at the interfaces or near the
interface regions, as recently done transport measurements
contacting the edges of these interfaces indicate \cite{ana12}.
The resolution of our TEM does not allow us to get more
information on the internal lattice structure within the interface
regions. Further studies using HRTEM are necessary to check
whether rhombohedral unit cells are embedded at the interfaces, a
not simple task indeed. We note however, that the irreversible
changes in the hysteretic response, after a temperature annealing
of less than one hour below 600~K already suggests that not only
the structure behind the interfaces but lattice defects and/or
hydrogen may play a role in the observed behavior.

\begin{figure}[]
\begin{center}
\includegraphics[width=.8\columnwidth]{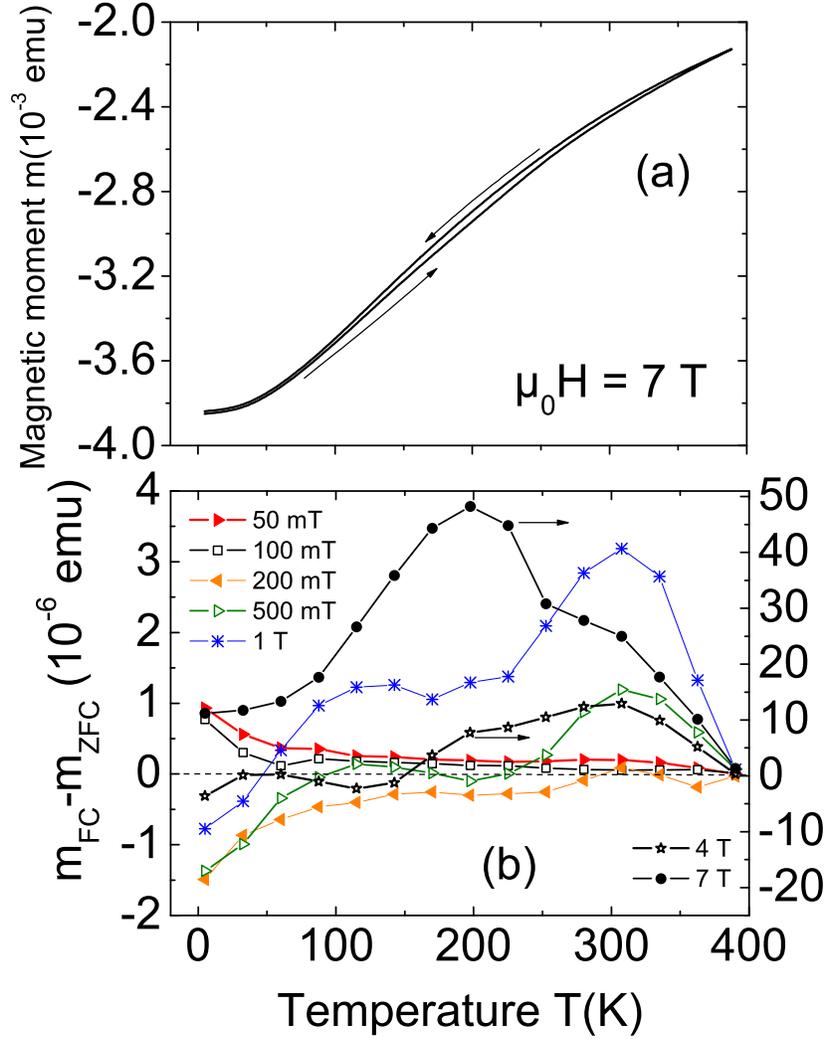}
\caption[]{(a) Magnetic moment $m$ vs. temperature at a constant
field of $\mu_0H= 7~$T in ZFC (lower curve) and FC (upper curve)
for sample HOPG-1. The field is applied normal to the graphene
layers and interfaces of the sample. (b) The difference between FC
and ZFC magnetic moment curves at different applied fields for
sample HOPG-1. Right $y-$axis corresponds to the fields of 4~T and
7~T.} \label{fc1}
\end{center}
\end{figure}

\section{Magnetization Results}
\subsection{Irreversible behavior in temperature at fixed fields}

A method that it does not need any background subtraction and
allows us to check for an intrinsic irreversibility due to either
pinned superconducting fluxons (or vortices) or magnetic domains
or magnetocrystalline anisotropy as in the case of ferromagnets,
is the measurement of the hysteresis between the magnetic moment
$m(T)$ in the FC and ZFC states. The opening between the ZFC- and
FC-branches at a field of 7~T applied normal to the graphene
layers and interfaces of the HOPG-1 sample can be clearly seen
already in the scale of Fig.~\ref{fc1}(a). The difference between
the FC- and ZFC-branches at different applied fields is shown for
this sample in Fig.~\ref{fc1}(b).

For a field of 50~mT  the difference between the FC- and
ZFC-branches $\Delta m = m_{\rm FC} - m_{\rm ZFC}$ is similar to
that obtained for the WTGP in ~\cite{sch12}, in particular the
shallow maximum near the turning temperature point (390~K) and the
decrease of this difference at 100~mT in the whole temperature
range. However, at 200~mT  $\Delta m$ is negative at nearly all
temperatures. Such a negative difference has been already seen in
WTGP but at temperatures only near the turning temperature point
(300~K) and was partially attributed to the influence of flux
creep at those high temperatures \cite{sch12}. However, in the
HOPG-1 sample the negative difference even increases at lower
temperatures, see Fig.~\ref{fc1}(b). This would indicate that the
possible superconducting regions shield or expel the applied field
stronger (more diamagnetism) when the sample is cooled down in
field than when warming in the ZFC state, clearly an unusual
behavior in conventional superconductors.

At higher applied fields the magnitude of the maximum $\Delta m(T
\simeq 310~$K), near the turning temperature point, increases as
well as the whole difference in all the temperature range with
exception of the low temperature region where the crossing to
negative values is observed up to a field of 4~T, see
Fig.~\ref{fc1}(b). The position of the high-temperature maximum
remains field independent, see Fig.~\ref{max}. A close look at the
$\Delta m(T)$ curves at fields $\mu_0 H \ge 200~$mT reveals that
maxima at $T \sim 150~$K and at $T \simeq 200~$K develop
increasing field.
\begin{figure}
\begin{center}
\includegraphics[width=.8\columnwidth]{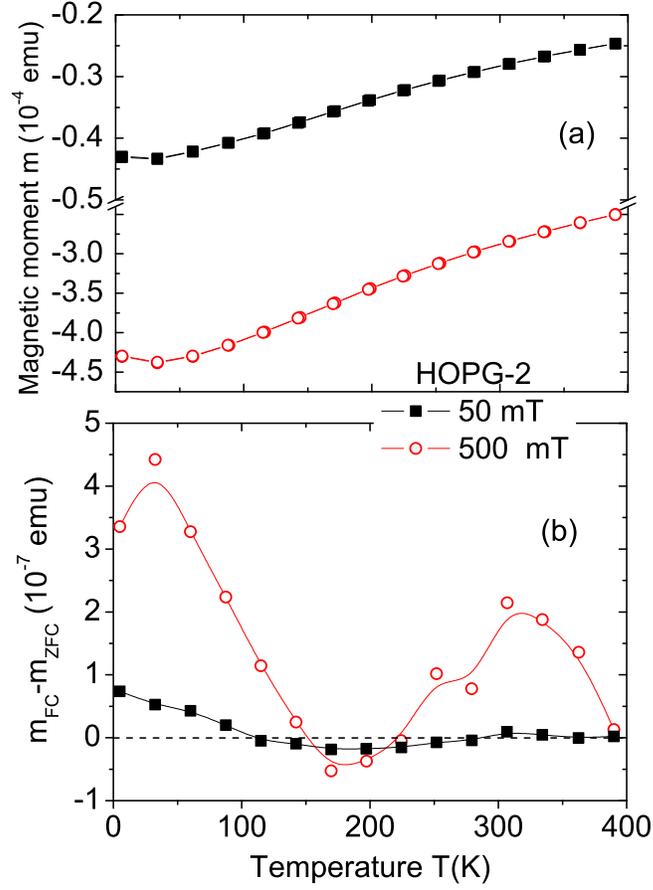}
\caption[]{(a) Temperature dependence of the magnetic moment for
sample HOPG-2. The curves were obtained in the ZFC (warming) and
FC (cooling) states, as for sample HOPG-1, see Fig.~\ref{fc1}. (b) The
difference $\Delta m(T)$  between the FC and ZFC curves at two applied fields. The
difference remains less than $1~\mu$emu in the
whole temperature range.} \label{fc2}
\end{center}
\end{figure}

To assure that the negative values as well as the whole
temperature dependence obtained for the difference between FC and
ZFC curves are not due to a SQUID artifact, the following
measurements were performed at 200~mT using two measurement mode
options, namely: (1) a linear regression mode and (2) the
iterative regression mode. It is known that the linear regression
mode can lead to smaller signals compared to the iterative
regression mode because only the amplitude of the signal is fitted
assuming the sample remains in the center point during the
complete measurement \cite{mce94}. Therefore, data were obtained
with the two methods to exclude errors due to a possible shift of
the sample during the measurements. The results (not shown) reveal
a small difference between the two mode curves, i.e.  the linear
regression mode curve is smaller than the iterative one with a
maximum difference of $0.2 \mu$emu, much smaller than the
differences observed, see Fig.~\ref{fc1}. This small difference
might be due to a displacement of the sample center position
determined by the regression mode.

To further rule out that the observed irreversibility in
temperature is not related to artifacts but is due to intrinsic
sample properties, especially at low fields $\mu_0 H < 1~$T where
this irreversibility is of the order of $\sim 1~\mu$emu (see
Fig.~\ref{fc1}(b)), the ZFC and FC curves have been measured also
for the HOPG-2 sample, see Fig.~\ref{fc2}. The obtained $\Delta
m(T)$ remains well below $1~\mu$emu at all temperatures, see
Fig.~\ref{fc2}(b). This result agrees with the absence of a field
hysteresis in sample HOPG-2, see Sec.~\ref{mag-f-h} below.

An opening of the ZFC-FC curves is usually associated with pinning
of magnetic entities starting at a field dependent irreversible
temperature. As noted in the measurements of the WTGP in
~\cite{sch12}, the  differences between FC and ZFC shown in
Fig.~\ref{fc1} for the HOPG-1 sample and for fields normal to the
interfaces do not appear to be compatible to the hysteresis one
expects from magnetic order of ferromagnetic large as well as
nanoparticles in the sample, see for example ~\cite{pro99}.
Furthermore, the amount of magnetic impurities in our samples is
too low to account for the measured hysteresis in temperature and
field and, more important, the irreversibility is observed only
for fields normal to the interfaces and not for fields applied
parallel to them.

Assuming that the existence of superconductivity is the reason for
the hysteresis in temperature, then  the observed differences in
the behavior of the FC-ZFC curves at different fields obtained for
the HOPG-1 sample -- relative to the those obtained for the WTGP
-- can be attributed to different superconducting phases. i.e.
phases with different critical temperatures $T_c$'s, or difference
pinning characteristics of the fluxons or vortices produced
between or within the Josephson-coupled grains. At the stage of
this research we are not able to provide any details on the
pinning characteristics necessary to understand the $\Delta
m(T,H)$ curves. Obviously, the pinning characteristics that these
curves suggest are not simple and a description of the overall
response has to take into account the Josephson coupling between
the grains and the response of the superconducting single domains
at fields above 40~mT, see Sec.~\ref{jos}. As a way to
characterize the existence of different $T_C$'s or the overall
pinning characteristics of the sample, we plot in Fig.~\ref{max}
the temperature of the maxima observed in $\Delta m(T,H)$ in
Fig.~\ref{fc1}. Interestingly, their amplitude but not their
position in temperature appears to depend on the applied field,
see Fig.~\ref{max}.

Experimental hints \cite{bar08,barint,sru11} as well as
recent evidence \cite{bal12} for the existence of embedded superconductivity at some
interfaces  in HOPG samples indicate that neither the critical temperature, nor the
temperature below which a Josephson coupling between
superconducting regions becomes effective, nor the distribution
in space of the superconducting phase(s) in the available HOPG samples is homogeneous. Therefore,
one may speculate that the observed maxima could be also related to regions
of the sample with different superconducting $T_c'$s.

\begin{figure}[t]
\begin{center}
\includegraphics[width=.80\columnwidth]{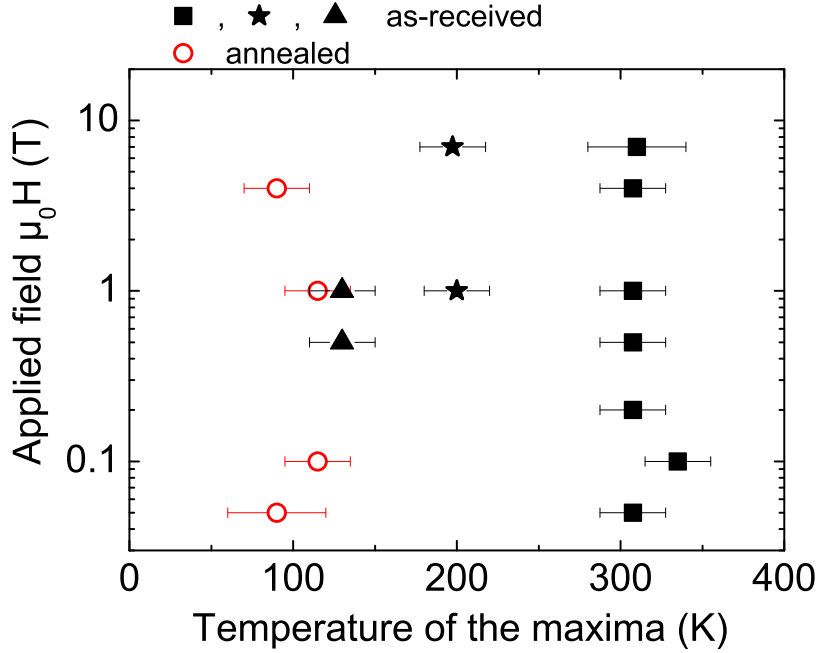}
\caption[]{Temperature of the maxima obtained from the difference
between FC and ZFC $\Delta m(T,H)$ curves vs. the applied field.
The close symbols correspond to the sample in the as-received
state and taken from Fig.~\ref{fc1} and the open symbol to the
same sample HOPG-1 after annealing, see Fig.~\ref{fc3} in
Sec.~\ref{anns}.} \label{max}
\end{center}
\end{figure}

Larger ZFC values (smaller in absolute value) than FC ones in the
magnetic moment are rather exotic and usually not observed,
neither in superconductors nor in ferromagnets and it is not known
to be a SQUID artifact. To our knowledge this anomalous behavior
in the FC-ZFC curves has been reported only for
quasi-two-dimensional Ru-based weak ferromagnetic superconductor
\cite{chi05} at one fixed magnetic field. The authors of that
paper suggested an interplay and coexistence of superconductivity
and weak ferromagnetic order as the origin for the anomaly but
without providing more detail how this may work. Qualitatively
speaking, we could speculate that this anomalous behavior is
related to a magnetic $p$-wave order parameter \cite{gon01} and
that the superconducting properties can be enhanced to some extent
under a magnetic field. In this case and upon the magnitude of the
vortex of vortex/fluxon pinning, the FC curve could expel more
field than the ZFC curve and a negative $\Delta m$ is possible.

\subsection{Magnetic field hysteresis}\label{mag-f-h}

\begin{figure}[t]
\begin{center}
\includegraphics[width=.8\columnwidth]{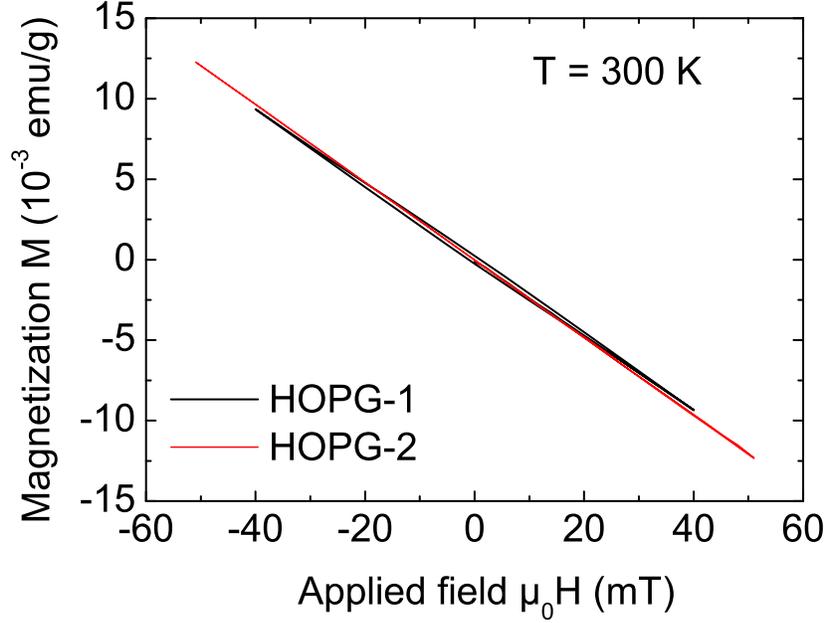}
\caption[]{Magnetization field loops for samples HOPG-1 and HOPG-2
at 300~K. The field was applied normal to the graphene layers. The
hysteresis is clearly seen in the scale of the figure for sample
HOPG-1, whereas sample HOPG-2 shows no hysteresis within
experimental error.} \label{dia}
\end{center}
\end{figure}

Field hysteresis loops for the two HOPG samples were measured at
different temperatures. The total response of the samples for
fields normal to the graphene layers is plotted in Fig.~\ref{dia}
at 300~K and for maximum applied fields of 40~mT and 50~mT for
samples HOPG-1 and HOPG-2, respectively. The measurements of the
field hysteresis start always at zero field after demagnetizing
the sample at 390~K and cooling down at zero field to the
measuring temperature. Within the scale of Fig.~\ref{dia} one can
recognize the field hysteresis for sample HOPG-1. The HOPG-2
sample shows no hysteresis within experimental resolution. Note
that both samples show nearly identical diamagnetic magnetization
values.  The hysteresis due to the superconducting solenoid used
by the SQUID apparatus produces a hysteresis artifact in the
magnetic moment signal, which is proportional to the diamagnetic
slope. This hysteresis is evident for fields above 100~mT in our
apparatus, see Fig.~4 in the supplementary information in
~\cite{sch12}. Taking also into account the larger mass of sample
HOPG-2 (larger absolute diamagnetic slope), the results shown in
Fig.~\ref{dia} indicate already that the field hysteresis found in
sample HOPG-1 is not a SQUID artifact but it is intrinsic of the
HOPG-1 sample.
\begin{figure}[t]
\begin{center}
\includegraphics[width=.8\columnwidth]{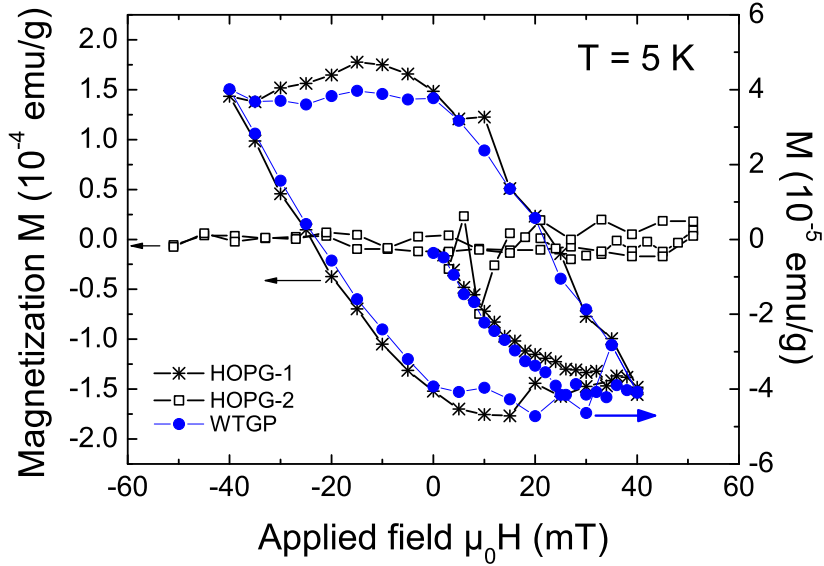}
\caption[]{Magnetic field loops of the magnetization for fields
applied normal to the graphene layers of the two HOPG samples
(left $y-$axis) and of the WTGP (sample S1 from ~\cite{sch12}) at
5~K. Linear diamagnetic backgrounds were subtracted from the
measured signals.} \label{hys1}
\end{center}
\end{figure}

\begin{figure}[t]
\begin{center}
\includegraphics[width=.75\columnwidth]{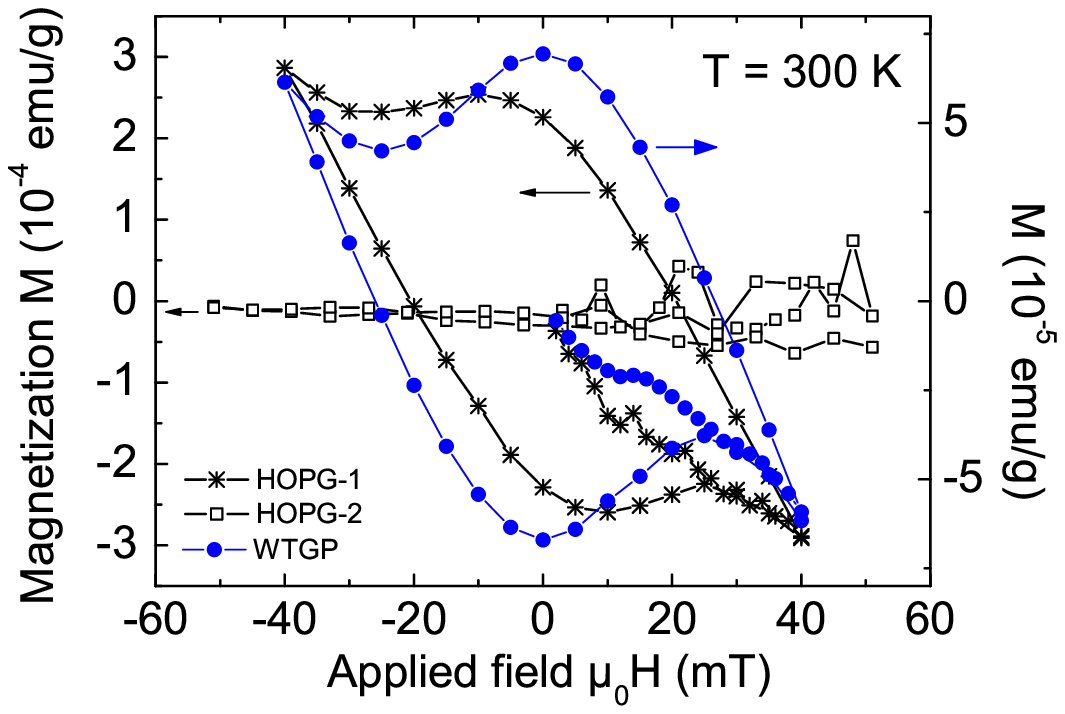}
\caption[]{The same as in Fig.~\ref{hys1} but at $T = 300~$K after
subtraction of the corresponding diamagnetic backgrounds, see
Fig.~\ref{dia}.} \label{hys2}
\end{center}
\end{figure}

After subtraction of the linear diamagnetic background ($m_D(5$K$)
= 3.62 \times 10^{-5}~$emu/gOe and $m_D(300$K$) = 2.26 \times
10^{-5}~$emu/gOe for sample HOPG-1) the obtained hysteresis are
plotted in Figs.~\ref{hys1} and \ref{hys2} for $T = 5~$K and
300~K.
For comparison the hysteresis of the WTGP \cite{sch12} were
included in the figures (right $y-$axis). The hysteresis for
samples HOPG-1 and WTGP (S1), look similar whereas the  HOPG-2
sample does not show any hysteresis at any temperature after
subtracting a similar diamagnetic slope as for sample HOPG-1.
\begin{figure}[t]
\begin{center}
\includegraphics[width=.75\columnwidth]{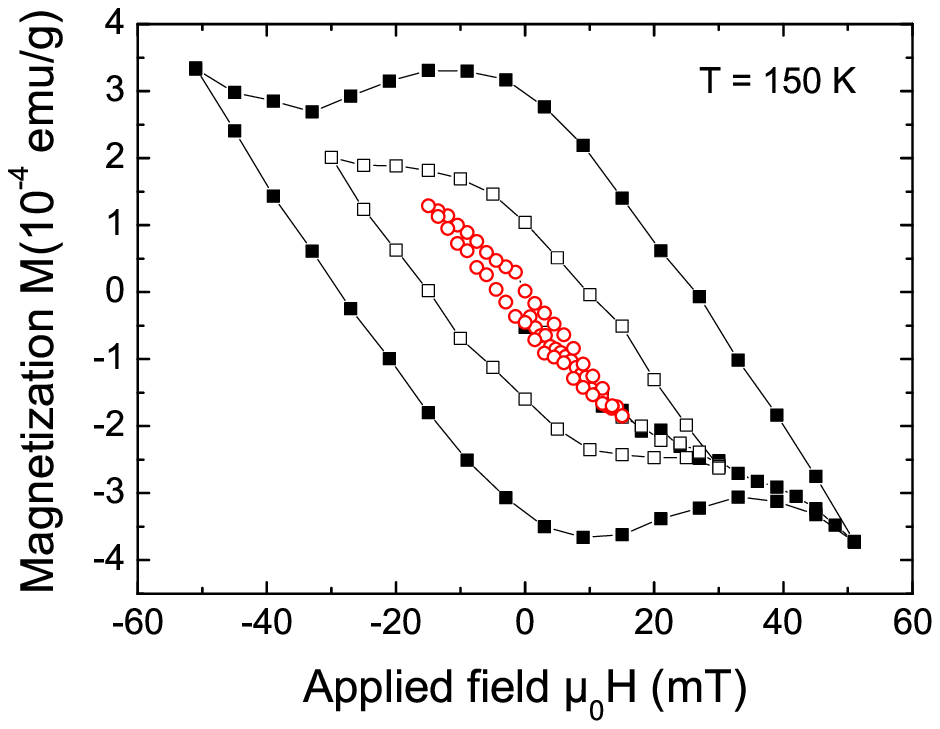}
\caption[]{Magnetic moment as a function of field applied normal
to the graphene planes and to the interfaces of sample HOPG-1 at
150~K. A constant diamagnetic slope was subtracted from the
measured data. Each loop has been measured for a given maximum
applied magnetic field.} \label{150}
\end{center}
\end{figure}
 The
obtained field hysteresis are superconductinglike and their  shape
depends on the maximum field applied as well as the temperature,
compare the loops at Figs.~\ref{hys1} and \ref{hys2}. The
hysteresis shape as well as its change with the maximum  field in
the loop are compatible with the existence of Josephson-coupled
grains inside the HOPG-1 sample, in agreement with the
current-voltage $I-V$ characteristic curves reported in
~\cite{bal12} for TEM lamellae obtained from similar HOPG samples.
\begin{figure*}[]
\begin{center}
\includegraphics[width=.85\columnwidth]{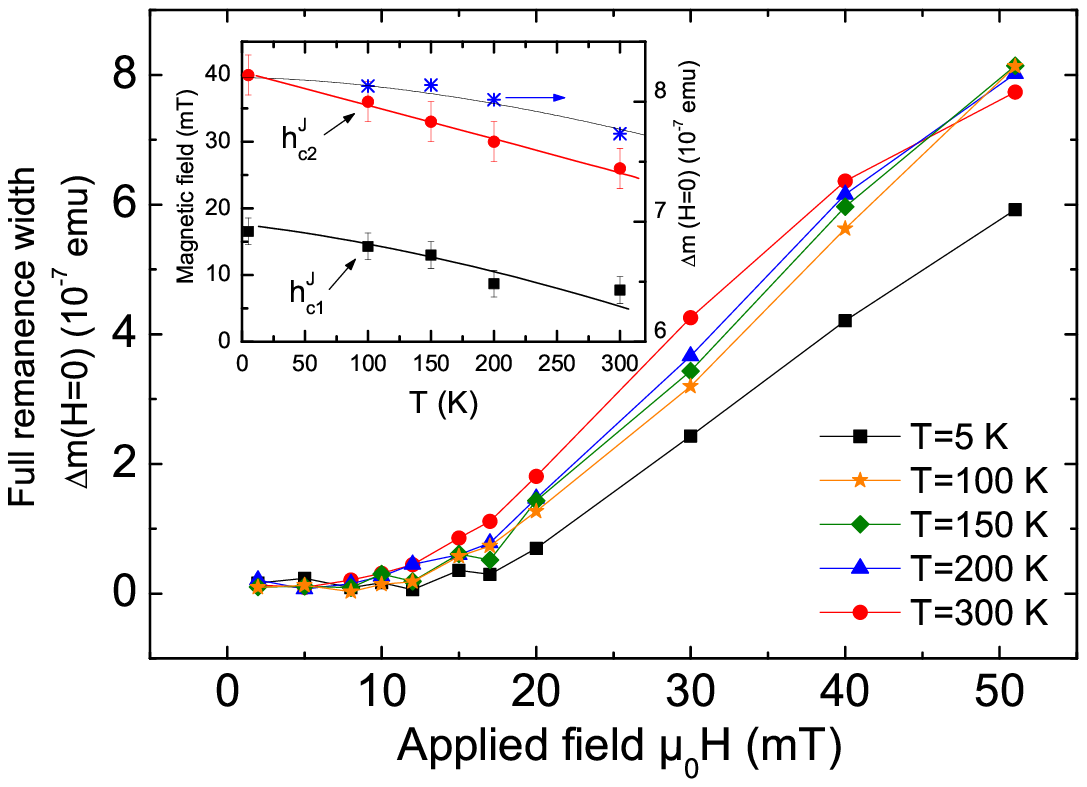}
\caption[]{Full remanence width   of the hysteresis loops $\Delta
m(H=0)$ measured after cycling the sample to a maximum field
$H_{\rm max}$ at different constant temperatures. The lower
Josephson critical field $h_{c1}^J$ shown in the insert (left
$y-$axis) is defined at the crossing point between the zero line
and a linear extrapolation of the increasing lines. Inset:
Temperature dependence of the lower (close squares) and upper
(close circles) critical Josephson fields (left $y$-axis). The
lines through the points are an aid to the eye. The right $y$-axis
corresponds to the full width of the hysteresis loop at zero field
after cycling the field to $H_{\rm max} = 50~$mT. The blue curve
follows $\Delta m(H=0,T) = 0.82[\mu$emu$] (1 -
(T[$K$]/1700)^{1.7})$.} \label{rem}
\end{center}
\end{figure*}

 Note that the superconductinglike
magnetization loops for the HOPG-1 sample, and for fields normal
to the embedded interfaces, are about three to four times larger
than the signals obtained for the WTGP sample. This should be not
surprising since, firstly, we obtain the magnetization values
dividing the superconductinglike magnetic moment by the total
sample mass not by the superconducting mass and therefore no
simple comparison can be really made. Second,
  due to the  huge anisotropy in the magnetic field
response  and because the interfaces are distributed randomly with respect to the
magnetic field in the case of the powder
sample  in contrast to the HOPG sample, we may expect
smaller superconductinglike signals in the powder sample.

As pointed out above, at fields applied parallel to the graphene
layers and to the interfaces there is no relevant hysteresis
between the FC and ZFC curves in temperature nor as a function of
applied field for both HOPG samples.  This qualitative change in
the magnetic response upon magnetic field direction has been
already reported for bulk HOPG samples in ~\cite{yakovjltp00}. The
results presented here confirm at least part of those results and
indicates clearly that the regions responsible for the measured
signals run parallel to the graphene layers of the graphite
sample. The clear two-dimensionality of the observed phenomenon
rules out the possibility that the observed irreversibility in
field or temperature is related to the existence of simple
ferromagnetism in our sample, intrinsic or due to impurities,
independently how small the ferromagnetic regions might be.

\subsection{Josephson critical fields}\label{jos}

As explained in ~\cite{sen91,bor91,and01,sch12} and due to the
granular superconductivity the magnetic field hysteresis of the
magnetization changes its shape upon the maximum magnetic field
applied at a given temperature. We expect to see a reversible,
hysteresis free behavior below a certain Josephson-critical field
$h_{c1}^J(T)$ to a Bean-like hysteresis at intermediate fields
below the upper critical Josephson field $h_{c2}^J(T)$. An
anomalous shrinking of the hysteresis plus a change in the slope
of the virgin curve are observed at fields above $h_{c2}^J(T)$
\cite{sen91,bor91,and01,sch12}. As an example, we show in
Fig.~\ref{150} the hysteresis loops at 150~K obtained at different
maximum applied fields for sample HOPG-1. The transition between a
non-hysteretic, reversible region to an irreversible one occurs at
$\mu_0h_{c1}^J(150$K$) \simeq 12~$mT, in agreement with the next
experiment described below.

The transition from reversible, zero remanence state to a finite
remanence can be also recognized by determining directly the full
 remanence width at zero-field. In this case the remanent loop width
at zero field  is defined as $\Delta m(H=0) = m_{H^+}(H=0) -
m_{H^-}(H=0)$, i.e. coming from the positive $H^+$ or negative
$H^-$ field branches. Figure~\ref{rem} shows $\Delta m(H=0)$ as a
function of $H_{\rm max}$ at different constant temperatures. The
whole behavior is similar to that measured for the WTGP
\cite{sch12} as well as in other high temperature superconductors
\cite{and01,mce90} and can be used to determine the Josephson
critical field $h_{c1}^J(T)$, which temperature dependence is
shown in the inset of this figure.

As in ~\cite{sch12} we define $h_{c2}^J(T)$  at the beginning of
the linear reversible region observed at high fields in the field
loop measurements. The so obtained $h_{c2}^J(T)$ is shown in the
inset of Fig.~\ref{rem}. Qualitatively speaking, the observed
behavior for sample HOPG-1, for fields normal to the interfaces,
is very similar to the one found for WTGP. The values of both
Josephson critical fields as well as their ratio are very similar
to those found for WTGP. However, the temperature dependence for
$h_{c1,c2}^J(T)$ does not appear to follow a logarithmic one, as
found for WTGP \cite{sch12}, but nearly a linear one. The
temperature dependence of the full remanence width at $H = 0$ and
after cycling the field to 50~mT is shown in the inset (right
$y-$axis). Though qualitatively similar, $\Delta m(H=0)(T)$
follows a slightly different temperature dependence as found for
the WTGP, indicating also a critical temperature clearly above
300~K.

\subsection{Annealing effects}\label{anns}

\begin{figure}[t]
\begin{center}
\includegraphics[width=.8\columnwidth]{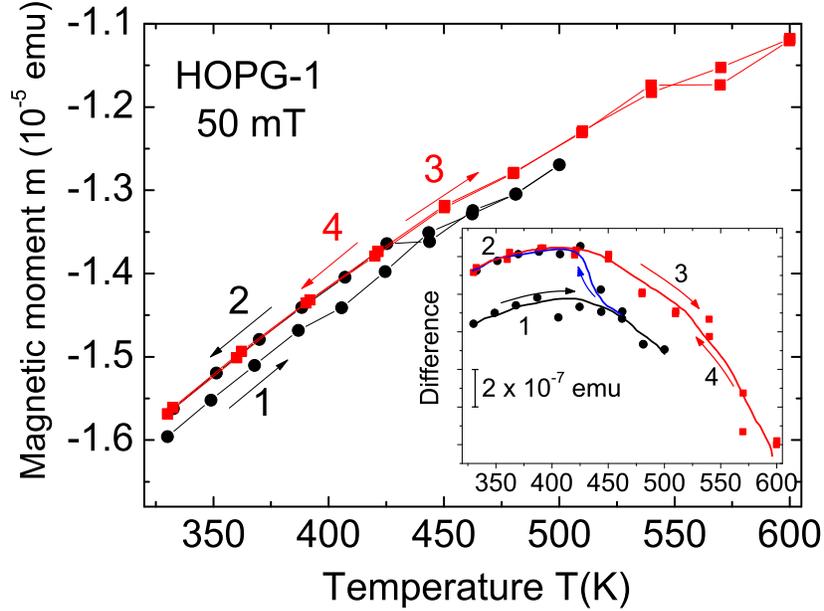}
\caption[]{Temperature dependence of the magnetic moment of sample
HOPG-1 at a field of 50~mT. The numbers and arrows indicate the
order and direction of the temperature sweeps. The inset shows the
same data as in the main panel but after subtraction of an
arbitrary background line given by the equation  $-1.12 \times
10^{-5} - 2\times 10^{-8} (600-T)$.} \label{ann}
\end{center}
\end{figure}

\begin{figure}[t]
\begin{center}
\includegraphics[width=.8\columnwidth]{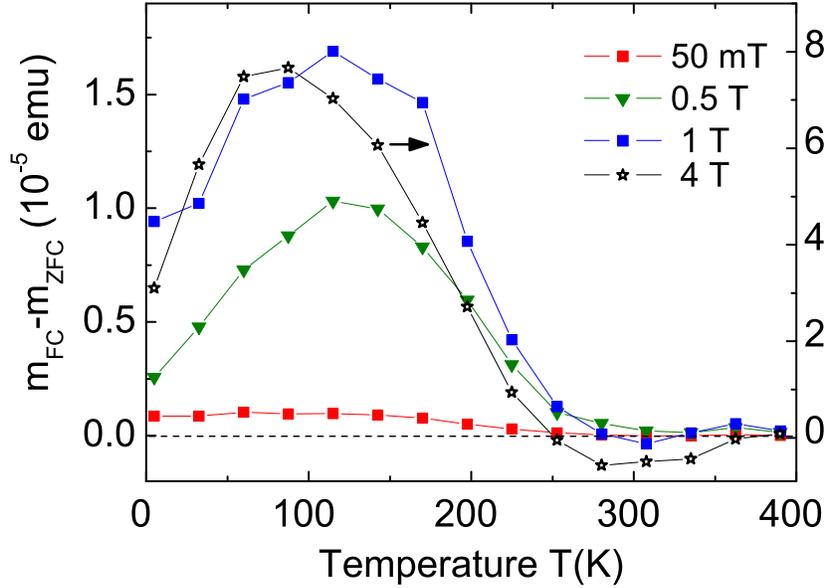}
\caption[]{Difference between the FC and ZFC curves measured for
sample HOPG-1 after annealing at different magnetic fields. The
right $y-$axis corresponds to an applied field of 4~T.}
\label{fc3}
\end{center}
\end{figure}

\begin{figure}[]
\begin{center}
\includegraphics[width=.7\columnwidth]{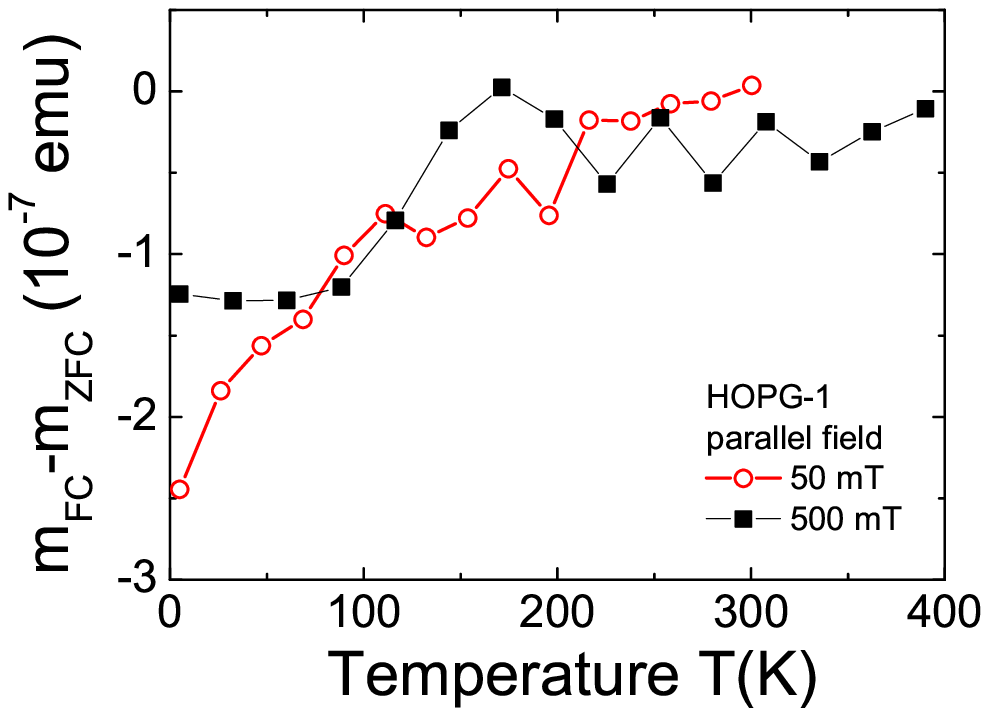}
\caption[]{Difference between the FC and ZFC curves measured for
sample HOPG-1 after annealing and for field applied parallel to
the graphene layers of the sample.} \label{par}
\end{center}
\end{figure}

After measuring the ZFC-FC and field hysteresis curves using the
maximum allowed temperature for the SQUID ($T < 400~$K) we
installed a SQUID oven, which enables measurements up to 800~K.
The HOPG-1 sample was fixed with graphite glue (which gives a
small diamagnetic contribution to the total signal) in a quartz
glass capillary to avoid any contact with the oven inner wall.
Before the measurements were started the oven was cleaned
thoroughly and heated up to 800~K (without sample) for one hour to
eliminate any possible magnetic contamination thereafter. The measurements
were performed starting at 330~K with the same procedure as for
the ZFC-FC curves described above. The measurements at all
temperatures in the SQUID oven were performed in helium
atmosphere.

In the first step and at 330~K we applied 50~mT field (always
normal to the internal interfaces) to the previously demagnetized
HOPG-1 sample and measured its magnetic moment up to 500~K
(step~1) and back to the initial temperature (step~2). We should
take into account that this kind of measurement, with a 25~K
temperature step, implies an annealing to the sample. In
particular, the sample was exposed to temperatures above 400~K for
nearly one hour till its temperature decreased below 450~K. After
the cycle $1 \rightarrow 2$, the procedure was repeated without
changing the applied field but up to a maximum temperature of
600~K and back (steps 3 and 4). The results are shown in
Fig.~\ref{ann}. In the inset of this figure we emphasize the
observed irreversibility after the first step, subtracting an
arbitrary background line. The opening of the curve between steps
1 and 2 can be clearly observed at $\sim 440$~K indicating that
part of the diamagnetic signal of the sample is lost after
annealing. The second cycle $3 \rightarrow 4$ with a maximum
temperature of 600~K shows a reversible behavior following the
same curve as that of step 2 below 425~K, see Fig.~\ref{ann}.

Figure~\ref{fc3} shows the difference between the FC and ZFC
curves after step 4 and after removing the  SQUID oven. The
observed irreversible behavior at different applied fields has to
be compared to the one obtained before annealing and shown in
Fig.~\ref{fc1}. It is clear that the annealing has produced a
clear change in the irreversible behavior of the sample. For
example, the irreversibility, which was observed at all fields
from the highest temperature turning  point of 390~K, see
Fig.~\ref{fc1}, is shifted clearly below 300~K after annealing,
see Fig.~\ref{fc3}. The maxima observed at $\sim 310~$K and $\sim
200~$K (in the as-received HOPG-1 sample) vanish after annealing,
leaving only an apparent single maximum at $\sim 100~$K, see
Figs.~\ref{fc3} and \ref{max}. We note also that although the
irreversibility temperature decreased below 300~K, the absolute
difference at the maximum increased by a large factor.

If we interpret the change in the whole irreversibility in
temperature after annealing as due to the vanishing or weakening
of some superconducting phase and/or change in the pinning
distribution or pinning strength, we may speculate on the
following possibilities. Annealing at 500~K might change either
the hydrogen concentration or simply the amount of defects at the
interfaces in which superconductivity is embedded. An annealing
temperature of 500~K appears to be high enough to release strains
and partially remove radiation damages in graphite \cite{pri78}.
On the other hand annealing can promote the formation of carbon
clusters or the grow of other lattice phases \cite{bol61}
affecting either the pinning properties or the critical
temperature. In case superconductivity would be solely related to
the existence of interfaces between Bernal and rhombohedral phases
\cite{kop12}, it appears highly unlikely that a temperature of
500~K or 600~K would affect them according to recently published
studies \cite{lui11}.

To check whether the annealing did any change to the magnetic
response of the sample for parallel fields, we have done ZFC and
FC measurements at two applied fields. The magnetic field was
applied parallel to the main area of the sample and parallel to
its graphene layers and interfaces within $\pm 3^\circ$.
Figure~\ref{par} shows the difference between the FC and ZFC
curves. The difference remains mostly within experimental error of
$\sim 10^{-7}~$emu, i.e. one to two orders of magnitude below the
one obtained for the perpendicular field direction, see
Fig.~\ref{fc3}.

\section{Conclusion}
In this work we studied the magnetic response of a HOPG sample
that shows similar irreversible behavior as the water treated
graphite powders reported recently. As for those samples, the
behavior observed in a HOPG sample in temperature and magnetic
field appears to be compatible with the existence of high
temperature granular superconductivity.  In agreement with
previous reports, this superconductinglike behavior is only
observed for field normal to the  internal interfaces (and to the
graphene planes) of the sample. Measurements in a similar HOPG
sample with similar diamagnetic response but without interfaces do
not show any irreversible behavior, neither in temperature nor as
a function of magnetic field, ruling out obvious SQUID artifacts.
Our results  support the view that the superconducting phase(s)
exist at or in embedded interfaces of the Bernal graphite matrix.
The clear two dimensionality of the observed phenomenon rules out
a ferromagnetic origin of the measured irreversibility. Because
these interfaces are not found in all HOPG samples, our study
provides an answer to the poor reproducibility of the
superconductinglike signals in HOPG samples. Temperatures between
400~K and 600~K can change irreversibly the superconductinglike
signals indicating that defects and/or hydrogen may play a role in
the observed behavior.

Acknowledgements: This work is supported by the Deutsche
Forschungsgemeinschaft under contract DFG ES 86/16-1.





\bibliographystyle{model3-num-names}


\end{document}